\documentclass[aps,prb,twocolumn,superscriptaddress]{revtex4-2}

\usepackage{graphicx}
\usepackage{bm}
\usepackage{color}
\usepackage{here}
\usepackage{lipsum}
\usepackage{amsmath}

\begin{document}

\title{Dynamics of orbital degrees of freedom probed via isotope $^{121,123}$Sb nuclear quadrupole moments in Sb-substituted iron-pnictide superconductors}

\author{T. Kouchi}\email[]{e-mail  address: takayoshi.kouchi@rs.tus.ac.jp}
\affiliation{Graduate School of Engineering Science, Osaka University, Osaka 560-8531, Japan}
\affiliation{Department of Applied Physics, Tokyo University of Science, Tokyo 125-8585, Japan}
\author{K. Yoshinaga}
\author{T. Asano}
\affiliation{Graduate School of Engineering Science, Osaka University, Osaka 560-8531, Japan}
\author{S. Nishioka}
\affiliation{Graduate School of Engineering Science, Osaka University, Osaka 560-8531, Japan}
\affiliation{National Institute for Materials Science, Ibaraki 305-0003, Japan}
\author{M. Yashima}
\author{H. Mukuda}\email[]{e-mail  address: mukuda.hidekazu.es@osaka-u.ac.jp}
\affiliation{Graduate School of Engineering Science, Osaka University, Osaka 560-8531, Japan}
\author{A. Iyo}
\affiliation{National Institute of Advanced Industrial Science and Technology (AIST), Ibaraki 305-8568, Japan}
\author{T. Kawashima}
\author{S. Miyasaka}
\affiliation{Graduate School of Science, Osaka University, Osaka 560-0043, Japan}

\date{\today}

\begin{abstract}
 
Isotope $^{121,123}$Sb nuclei with large electric quadrupole moments are applied to investigate the dynamics of orbital degrees of freedom in Sb-substituted iron(Fe)-based compounds. 
In the parent compound LaFe(As$_{0.6}$Sb$_{0.4}$)O, the nuclear spin relaxation rate $^{121,123}(T_{1}^{-1})$ at $^{121,123}$Sb sites was enhanced at structural transition temperature ($T_{s}\sim$ 135 K), which is higher than N\'eel temperature ($T_{\rm N}\sim$125 K). 
The isotope ratio $^{123}(T_{1}^{-1})/^{121}(T_{1}^{-1})$ indicates that the electric quadrupole relaxation due to the dynamical electric field gradient at Sb site increases significantly toward $T_{s}$.
It is attributed to the critically enhanced nematic fluctuations of stripe-type arrangement of Fe-$3d_{xz}$ (or $3d_{yz}$) orbitals. 
In the lightly electron-doped superconducting (SC) compound LaFe(As$_{0.7}$Sb$_{0.3}$)(O$_{0.9}$F$_{0.1}$), the nematic fluctuations are largely suppressed in comparison with the case of the parent compound, however, it remains a small enhancement below 80 K down to the $T_c$($\sim$ 20 K). 
The results indicate that the fluctuations from both the spin and orbital degrees of freedom on the $3d_{xz}$(or $3d_{yz}$) orbitals can be seen in lightly electron-doped SC state of LaFeAsO-based compounds.
We emphasize that isotope $^{121,123}$Sb quadrupole moments are sensitive local probe to identify the dynamics of orbital degrees of freedom in Fe-pnictides, which provides with a new opportunity to discuss the microscopic correlation between the superconductivity and both nematic and spin fluctuations simultaneously even in the polycrystalline samples. 

\end{abstract}


\maketitle

\section{Introduction}

Iron(Fe)-based superconductivity LaFeAs(O$_{1-y}$F$_{y}$) ($T_c$= 26 K) appears in the vicinity of stripe-type antiferromagnetic (AFM) order and the structural(nematic) transition from tetragonal($C_4$) to orthorhombic($C_2$) phases\cite{Kamihara2008,H.Luetkens}, and hence the mechanism of the superconductivity have been discussed based on  spin fluctuations and nematic(orbital) fluctuations\cite{I.I.Mazin2008,K.Kuroki2008,H.Kontani2010,S.Onari2012}.
In most Fe-based superconductors composed of hole and electron Fermi surfaces (FSs) of similar size, it has been reported that the low energy spin fluctuations enhanced toward low temperatures have some correlation with $T_c$\cite{T.Imai2009,Ning2010,Y.Nakai2010,T.Oka2012,T.Shiota2016,P.Wiecki2018}. 
Simultaneously, nematic fluctuations critically-enhanced in the SC phase has been reported in single-crystals of some Fe-based families, which also revealed a relation with the SC phases\cite{M.Yoshizawa2012,Y.Gallais2013,A.E.Bohmer2014,H.Hosoi2016}.
In some of the Fe-based SC families,  the further high-$T_c$ phases appear again in heavily-electron-doped regimes without the nested FSs such as in LaFeAsO$_{1-y}$(F/H)$_{y}$\cite{Iimura2012,M.Hiraishi2014} and intercalated FeSe\cite{Guo,Y.Mizuguchi2011,A.K.Maziopa2011}: In those compounds, the spin fluctuations at low energies are not critically enhanced in contrast to the former cases in lightly electron-doped SC states. 
Recently, the analyses in collaboration with the Knight shift measurement enabled us to extract the characteristic behavior of spin fluctuations with gap at low energies, which become prominent in the re-enhanced high-$T_c$ state of the heavily electron-doped regimes\cite{T.Kouchi2022}, and hence it is proposed to be one of the indispensable factors from the viewpoint of ``spin''-based scenario. 
On the one hand, due to the lack of large single crystals, the contribution of orbital degrees of freedoms is still unclear in LaFeAsO(La1111)-based compounds.
The universality and diversity of  SC mechanism should be identified in Fe-based compounds over a wide doping region, including the roles of orbital degrees of freedom in addition to the roles of spin components.

To shed light on the dynamical features of orbital degrees of freedom, we focus on the Sb-substituted Fe-pnictides LaFe(As,Sb)(O,F) in this study\cite{Carlsson2011,T.Kawashima2021}. 
In general, the nucleus with $I\ge1$ possess not only a magnetic moment but also an electric quadrupole moment($Q$). 
As shown in Table \ref{nuclei}, the Sb nucleus possesses a relatively large $Q$ that could interact with the possible orbital dynamics through the local fluctuations of the electric field gradient (EFG) at Sb site. 
Unfortunately, the earlier $^{75}$As($I$=3/2)-NMR studies in the La1111-based compounds\cite{Y.Nakai2008,N.Terasaki2009,H.J.Grafe2008} could not capture the dynamics of the orbital fluctuations, due to the lack of sensitivity.
The comparison of isotope $^{121,123}$Sb nuclei ($I>$1) would give us a new insight into nematic fluctuations derived from the degenerate Fe-$3d$ orbitals through the local fluctuations of EFG at Sb site. 

In this paper, we report the first isotope $^{121,123}$Sb-NMR investigation on Sb-substituted Fe-pnictides LaFe(As$_{1-x}$Sb$_{x}$)(O$_{1-y}$F$_{y}$) in the parent and lightly electron-doped SC($\equiv$SC1) phases, detecting the temperature evolution of  the low-energy nematic fluctuations of stripe-type arrangement of Fe-$3d_{xz}$ (or $3d_{yz}$) orbitals. 
Our findings reveal that the isotope $^{121,123}$Sb quadrupole moments are very sensitive nuclear probe to identify the dynamics of orbital degrees of freedom in Fe-pnictides, which provide with the opportunity to discuss the nematic fluctuations in addition to spin fluctuations systematically in the various types of Fe-pnictides.

\begin{figure}[htbp]
\centering
\includegraphics[width=7.5cm]{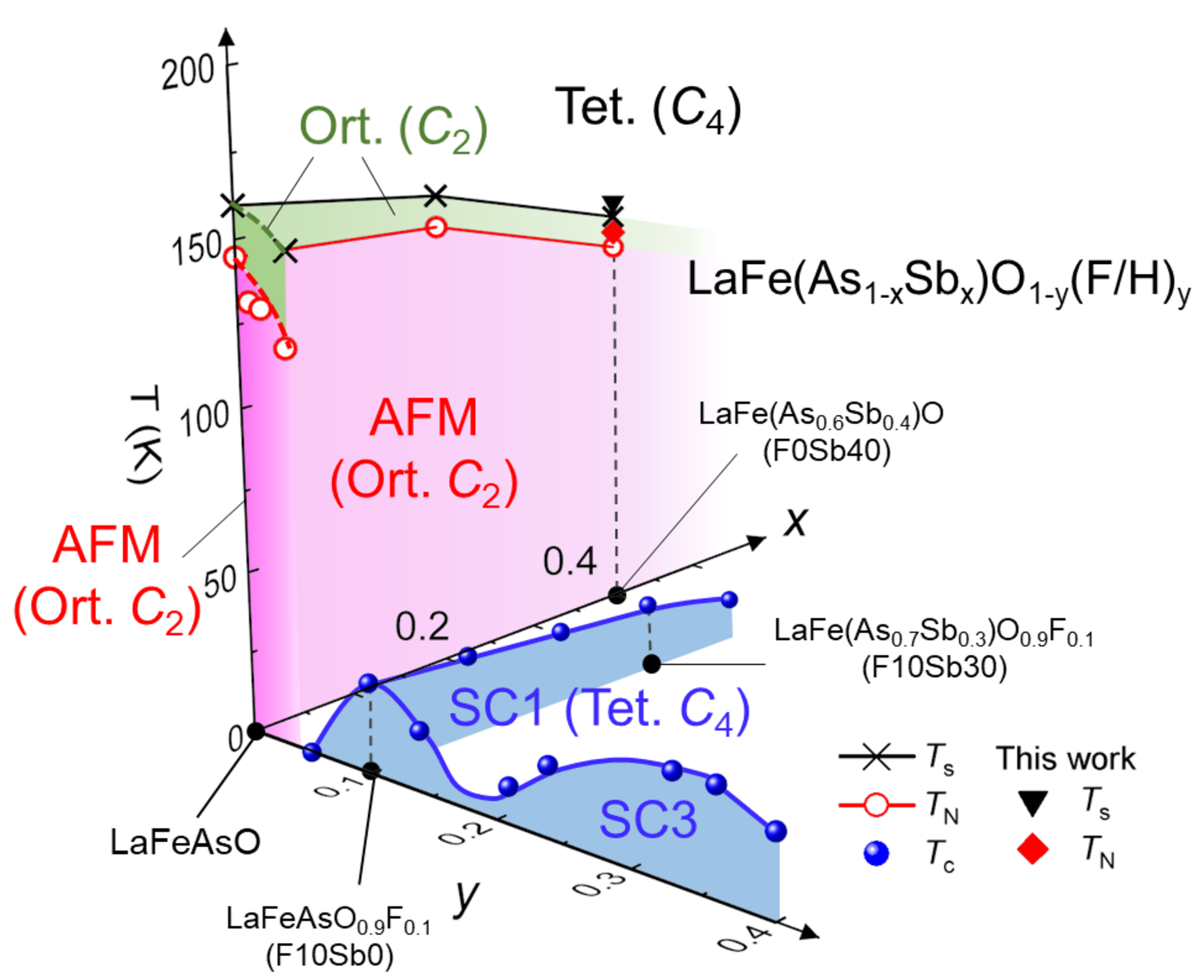}
\caption[]{(Color online)  
Phase diagram of LaFe(As$_{1-x}$Sb$_{x}$)O$_{1-y}$(F/H)$_{y}$ obtained from the Refs.\cite{Carlsson2011,Iimura2012,T.Kawashima2021}.
The present NMR studies are performed on the parent LaFe(As$_{0.6}$Sb$_{0.4}$)O (Denoted as ``F0Sb40''), and the lightly electron-doped SC1 compounds  LaFe(As$_{0.7}$Sb$_{0.3}$)(O$_{0.9}$F$_{0.1}$) (``F10Sb30'', $T_c$=20 K), together with the Sb-free LaFeAs(O$_{0.9}$F$_{0.1}$) (``F10Sb0'', $T_c$= 28 K) for comparison.
}
\label{T1}
\end{figure}

\begin{table}[h]
\caption{Nuclear spin ($I$), gyromagnetic ratio ($\gamma_{\rm n}$), and quadrupole moment($Q$) for nuclei in the La1111-based compounds\cite{Abragam,Slichter,S.Kitagawa2013}.}
\label{table1}
\begin{tabular}{@{\hspace{0.5cm}}c@{\hspace{0.5cm}}@{\hspace{0.4cm}}c@{\hspace{0.4cm}}c@{\hspace{0.4cm}}c}
\centering

& $I$ & $\gamma_{\rm n}$ (MHz/T) & $Q$ (10$^{-24}$cm$^{2}$) \\
\hline
$^{57}$Fe & 1/2 & 1.3785 & -- \\
$^{75}$As & 3/2 & 7.292 & $+$0.29 \\
$^{139}$La & 7/2 & 6.0142 & $+$0.2 \\
\hline
$^{121}$Sb & 5/2 & 10.189 & $-$0.543 \\
$^{123}$Sb & 7/2 & 5.5175 & $-$0.692 \\

\label{nuclei}
\end{tabular}
\end{table}

\section{experimental}

Polycrystalline samples of LaFe(As$_{1-x}$Sb$_{x}$)(O$_{1-y}$F$_{y}$) were synthesized using a solid state reaction method\cite{T.Kawashima2021}. 
Powder x-ray diffraction (XRD) measurement indicates that the lattice parameters monotonically change with $x$ and $y$, as reported in Ref.\cite{T.Kawashima2021}.
The $T_c$ values were determined from an onset of zero-resistivity and diamagnetic response in $dc$ susceptibility measurement\cite{T.Kawashima2021}.
NMR measurements  are performed at $^{121,123}$Sb, $^{75}$As, and $^{139}$La sites for the coarse-powder samples of Sb-substituted parent LaFe(As$_{0.6}$Sb$_{0.4})$O (denoted as ``F0Sb40'' hereafter), and lightly electron-doped SC compounds LaFe(As$_{0.7}$Sb$_{0.3}$)(O$_{0.9}$F$_{0.1}$) (``F10Sb30'', $T_c$=20 K) and LaFeAs(O$_{0.9}$F$_{0.1}$) (``F10Sb0'', $T_c$= 28 K).
The nuclear spin-lattice relaxation rates ($1/T_{1}$)  were measured at the resonance peak corresponding to the external field
perpendicular to the c axis denoted by the allows in Figs. 2(a)-(c), and were determined by fitting a recovery curve for nuclear magnetization to the multiple exponential functions,  $m(t)=0.1\exp(-t/T_{1})+0.9\exp(-6t/T_{1})$ for $I$=3/2, $m(t)=0.028\exp(-t/T_{1})+0.178\exp(-6t/T_{1})+0.794\exp(-15t/T_{1})$ for $I$=5/2, and $m(t)=0.012\exp(-t/T_{1})+0.068\exp(-6t/T_{1})+0.206\exp(-15t/T_{1})+0.714\exp(-28t/T_{1})$ for $I$=7/2\cite{Narath}.

\begin{figure}[htbp]
\centering
\includegraphics[width=9.5cm]{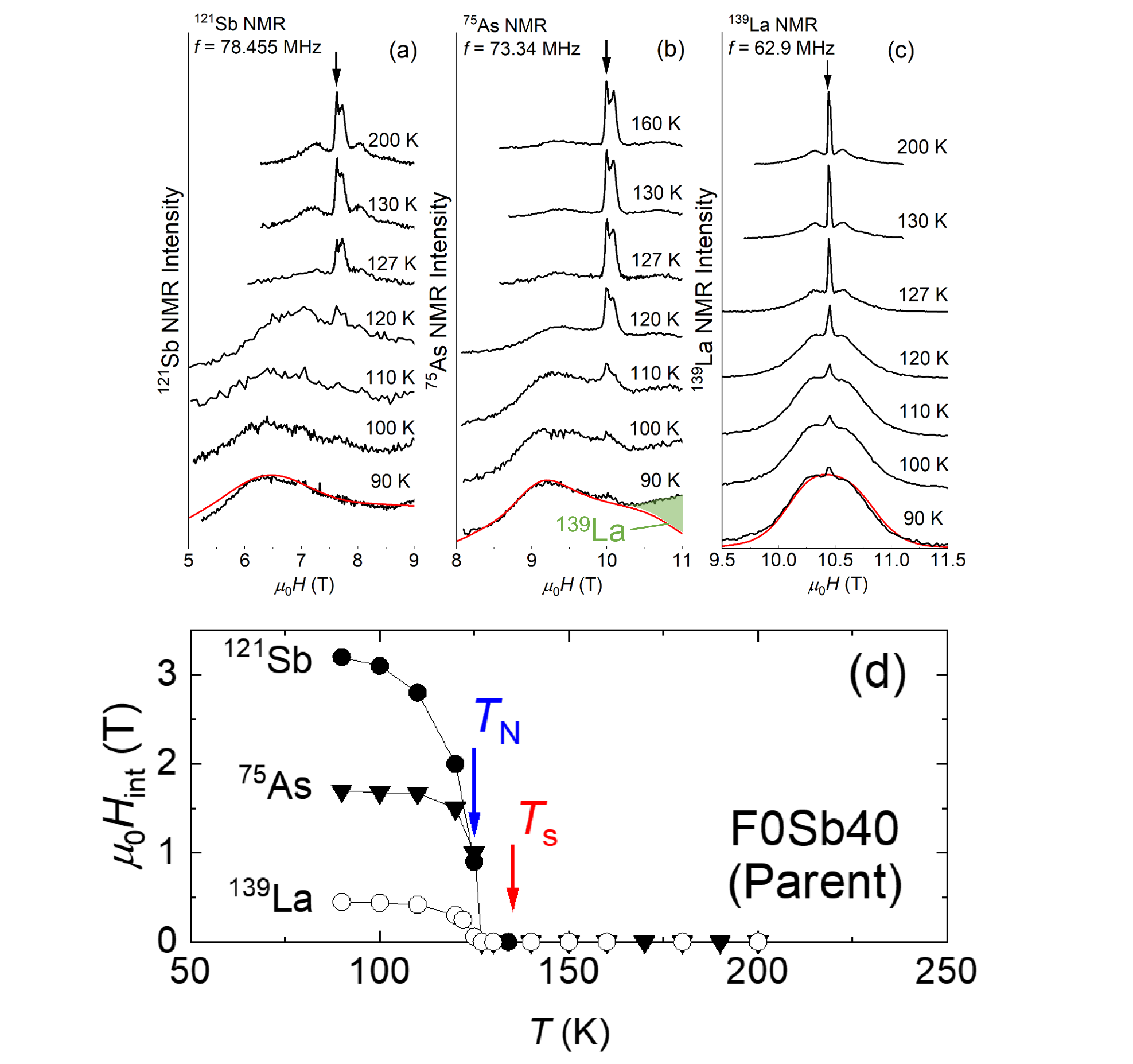}
\caption[]{(Color online) 
$T$ dependence of (a)$^{121}$Sb, (b)$^{75}$As, and  (c)$^{139}$La-NMR spectra for  the parent ``F0Sb40''. The allows denote the resonance peak corresponding to $H\perp c$.
Solid curves at 90 K($\ll T_{\rm N}$) are simulations obtained by assuming the $H_{\rm int}$ and $\nu_{\rm Q}$ at each site(see text). 
(d) $T$ dependence of $H_{\rm int}$ for all nuclear sites indicates the anomaly due to the static AFM order below $T_{\rm N}\sim$ 125 K for F0Sb40, which is lower than the structural transition at $T_{s}\sim$135 K.}
\label{phase diagram}
\end{figure}

\section{Results and Discussion}

\subsection{Parent compound LaFe(As$_{0.6}$Sb$_{0.4})$O (``F0Sb40'')}

Figure \ref{phase diagram}(a) shows the temperature ($T$) dependence of the $^{121}$Sb-NMR spectra for the parent ``F0Sb40''.
The typical powder pattern spectra are observed at high temperatures.
The spectra become broader upon cooling below 125 K, which is also seen in  $^{75}$As- and $^{139}$La-NMR spectra, as shown in Figs. \ref{phase diagram}(b) and \ref{phase diagram}(c), respectively. 
As shown by the solid curves in these figures, the NMR spectra at 90 K can be reproduced by the parameters of the internal field ($H_{\rm int}$) and NQR frequency ($\nu_{\rm Q}$) for each nuclear site: These are  $^{121}H_{\rm int}\sim$3.2 T and $^{121}\nu_{\rm Q}\sim$10.5 MHz for $^{121}$Sb site,  $^{75}H_{\rm int}\sim 1.6$ T and  $^{75}\nu_{\rm Q}\sim$ 11.5 MHz for $^{75}$As site, and $^{139}H_{\rm int}\sim 0.34$ T and $^{139}\nu_{\rm Q}\sim$ 1.8 MHz for $^{139}$La site. 
These $H_{\rm int}$ are comparable to those of typical parent iron-pnictides in the stripe AFM order, such as BaFe$_{2}$As$_{2}$\cite{K.Kitagawa2008} and LaFeAsO\cite{N.Terasaki2009}.
The $H_{\rm int}$ for these nuclear sites exhibit an rapid increase below 125 K, as summarized in Fig. \ref{phase diagram}(d), indicating that the AFM order develops below N\'eel temperature $T_{\rm N}\sim$ 125 K  homogeneously over the crystals.
The values of $H_{\rm int}$ depend on that of the hyperfine coupling constant ($^{i}\!\!A_{\rm hf}$) at these nuclear sites, since it is given by the relation $^{i}H_{\rm int}= ^{i}\!\!\!A_{\rm hf}M_{\rm AFM}$, where $M_{\rm AFM}$ is an AFM moment at the Fe site. 
By using $^{75}A_{\rm hf}\sim 2.0$-$2.5$T/$\mu_{\rm B}$ estimated at the As site of  La1111\cite{H.J.Grafe2008,F.Sakano2019}, the $M_{\rm AFM}$ can be estimated to be $0.64$-0.80 $\mu_{\rm B}$, which is comparable to the values observed for the parent LaFeAsO\cite{N.Qureshi2010}.
We note that the $^{121}H_{\rm int}$ at Sb site was the largest among the values at these nuclei,  owing to the largest magnetic hyperfine coupling between the Sb nucleus and the Fe-$3d$ electrons. It is due to that the $5p$ orbitals of the Sb are more extended than that of the As. The relative ratios were evaluated to be $^{121}A_{\rm hf}/^{75}A_{\rm hf}\sim2$ and $^{121}A_{\rm hf}/^{139}A_{\rm hf}\sim9.4$ in this experiment.

\begin{figure}[htbp]
\centering
\includegraphics[width=8.5cm]{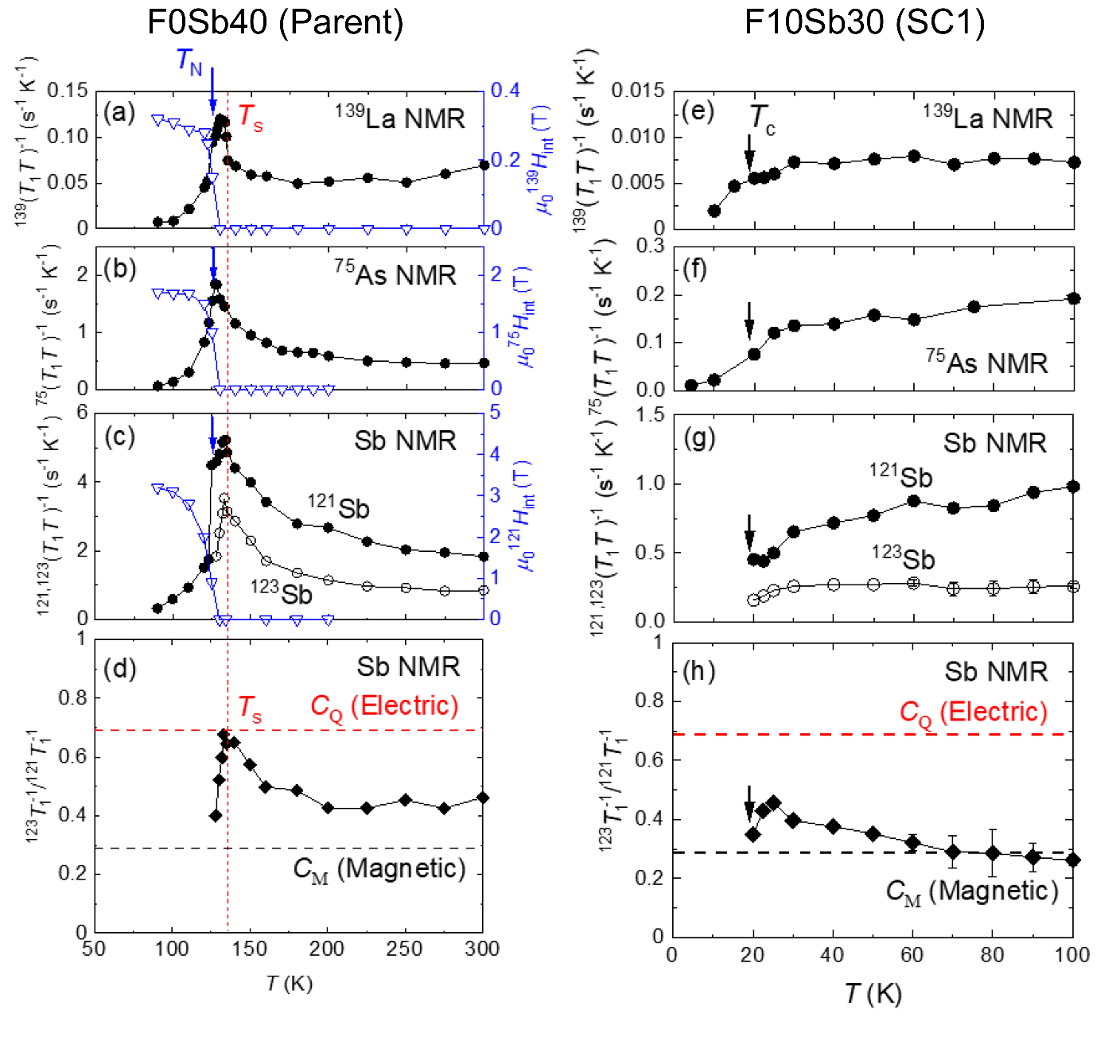}
\caption[]{(Color online)  
$T$ dependence of $1/T_{1}T$ and the $H_{\rm int}$ at (a) La, (b)As, and (c) Sb sites of the parent (F0Sb40). 
The $1/T_{1}T$s at three nuclear sites increase significantly upon cooling toward $T_{\rm N}\sim125$ K, while only $1/T_{1}T$ at Sb site shows an additional anomaly at $T_s\sim$135 K. 
(d) $T$ dependence of the ratio $^{123}(T_{1}^{-1})/^{121}(T_{1}^{-1})$ at isotopic Sb sites for F0Sb40, indicating that the electric relaxation becomes dominant around $T_{s}$ due to the critical slowing down of the orbital fluctuations. 
As for the lightly doped SC sample (F10Sb30), the results of $1/T_{1}T$ for (e) La, (f)As, and (g) Sb sites are shown. (h) Their isotope ratio $^{123}(T_{1}^{-1})/^{121}(T_{1}^{-1})$ suggests the moderate enhancement due to the nematic fluctuations below 80 K in the lightly doped SC1 phase.
}
\label{Sb_T1_ref}
\end{figure}

Next, we focus on the electronic states through the measurement of the nuclear spin relaxation rate.
In the parent F0Sb40, the $1/T_{1}T$ increases upon cooling and exhibits a peak at $T_{\rm N}\sim125$ K due to the static AFM order, as shown in Figs. \ref{Sb_T1_ref}(a)-\ref{Sb_T1_ref}(c). 
Remarkably,  additional peak at 135 K observed for Sb-NMR probe corresponds to the structural (nematic) transition at $T_s$ that appears slightly above $T_{\rm N}$ (See Fig. \ref{Sb_T1_ref}(c)).  
It suggests that the other relaxation mechanisms are added at Sb nucleus, which is observed more sensitively than at the other nuclei.
In general, when the nucleus possesses an electric quadrupole moment ($Q$), the observed relaxation rate are composed of the magnetic contribution $(1/T_{1}T)_{\rm M}$ and electric one $(1/T_{1}T)_{\rm Q}$. The $(1/T_{1}T)_{\rm M}$ is generally described as, 
\begin{equation}
\label{equ1}
\left(\frac{1}{T_{1}T}\right)_{\rm M} \propto \gamma^{2}_{\rm n} \lim\limits_{\omega \to 0}\sum\limits_{\vec q}A_{\rm hf}(\vec q)^{2}\frac{\chi_{\rm m}^{\prime\prime}(\vec q,\omega)}{\omega}
\end{equation}
where $A_{\rm hf}(\vec q)$ is the hyperfine-coupling constant at $\vec q$, and $\chi_{\rm m}^{\prime\prime}(\vec q,\omega)$ is dynamical spin susceptibility at wave vector $\vec q$ and energy $\omega$\cite{Abragam,Slichter}.
In usual metals and alloys,  the magnetic part $(1/T_1T)_{\rm M}$ is predominant  through the large Fermi-contact interaction with electron spins on conduction bands, resulting in that the electric relaxation part $(1/T_1T)_{\rm Q}$ is negligible. 
However, in the case of Fe-pnictides, the electric part $(1/T_{1}T)_{\rm Q}$ is possible to be enhanced by the fluctuations of EFG between the $C_2$ and $C_4$ symmetry.
According to the previous studies\cite{Suter,A.Suter2000,A.P Dioguardi2016,YObata1,YObata2,I.Vinograd2022}, it could be expressed by,
\begin{equation}
\label{equ2}
\left(\frac{1}{T_{1}T}\right)_{\rm Q} \propto f(I)(eQ)^{2}\lim\limits_{\omega \to 0}\sum\limits_{\vec q}\frac{\chi_{\rm nem}^{\prime\prime}(\vec q,\omega)}{\omega}
\end{equation}
where the $f(I)$ is $(2I+3)/(I^{2}(2I-1))$, and $\chi_{\rm nem}^{\prime\prime}$ is the dynamical nematic susceptibility\cite{A.P Dioguardi2016}.
Thus, if only magnetic relaxation mechanism is dominant, the relaxation rate should be proportional to the square of $\gamma_{\rm n}$, resulting in that an isotopic ratio of the relaxation rate $(^{123}T_{1}^{-1})_{\rm M}/(^{121}T_{1}^{-1})_{\rm M}$ approaches ($^{123}\gamma_{\rm n}/^{121}\gamma_{\rm n})^2 = 0.293(\equiv C_{\rm M})$. 
In contrast, if only electric relaxation mechanism is dominant, the isotopic ratio $(^{123}T_{1}^{-1})_{\rm Q}/(^{121}T_{1}^{-1})_{\rm Q}$ is expected to be $f(7/2)(^{123}Q^{2})/f(5/2)(^{121}Q^{2})=0.691(\equiv C_{\rm Q})$.

Figure \ref{Sb_T1_ref}(d) shows the $T$ dependence of $^{123}(T_{1}^{-1})/^{121}(T_{1}^{-1})$. 
Remarkably, it is close to the value of $C_{\rm Q}$ at around $T_s$, indicating that the electric relaxation process at Sb site is critically enhanced toward $T_s$. 
It is obviously attributed to the critical slowing down for the stripe-type alignment of the Fe-$3d_{xz}$ (or $3d_{yz}$) orbital at $T_s$ from the tetragonal($C_4$) to the orthorhombic($C_2$) phase. 
The fluctuations of in-plane anisotropy in EFG are induced by the dynamics of the local charge distribution at Sb-5$p_{x,y}$ orbitals that hybridize with Fe-3$d_{xz,yz}$ orbitals, as discussed in the lightly-doped 122 based Fe-pnictides\cite{A.P Dioguardi2016}.
We also note that, at high temperatures, this ratio stays at the intermediate value ($\sim0.45$) between $C_{\rm M}$ and $C_{\rm Q}$, indicating that the magnetic and electric relaxation components are almost comparable even above $T_s$. 
It suggests that the nematic fluctuations between the $C_4$ and $C_2$ symmetries remain even in the $C_4$ phase well above $T_{s}$. 

Next we attempt to extract the electric relaxation component to estimate the outline of $T$ evolution of nematic fluctuations. Here we made the approximation of simple addition of the relaxation rates of the two channels expressed as, $^{i}(1/T_{1}T)(\equiv\ \!\!^{i}R) =\ ^{i}(1/T_{1}T)_{\rm M} + \ ^{i}(1/T_{1}T)_{\rm Q} (i=121, 123)$. In this analysis, we should note that, when the $(1/T_{1}T)_{\rm Q}$ term is more enhanced, the functional form of the relaxation is a complex mixture of terms arising from both magnetic and quadrupole interaction channels\cite{Suter,A.Suter2000}.
We attempt to use the same functional form of the relaxation shown in the section I\hspace{-1.2pt}I, which will give us an information of the outline of $T$-evolution of nematic fluctuations over some ambiguity in the absolute value of $T_{1}$.
First, we consider that the orbital fluctuations also contribute to the magnetic relaxation very rarely through the possible terms, $(1/T_{1}T)_{\rm M}^{\rm orb} \propto \gamma_{\rm n}^{2}\lim\limits_{\omega \to 0}\sum_{\vec{q}}(A_{\rm L})^{2}\chi''_{\rm nem}(\vec{q}, \omega)/\omega$\cite{YObata1,YObata2}, where the $A_{\rm L}$ is the magnetic hyperfine coupling constant that appears due to the coupling with $\chi_{\rm nem}$ in the case of Fe-pnictides. 
It will cause the magnetic relaxation, however, it may be negligibly smaller than the component of spin fluctuations(eq.(1)), since the ratio $^{123}(T_{1}^{-1})/^{121}(T_{1}^{-1})$ is close to $C_{\rm Q}$ at $T_s$, where the $\chi_{\rm nem}$ is largely enhanced. 
Therefore, it is assumed that the magnetic(spin) and electric(orbital) relaxation components may be handled independently in the relaxation process. Thus, the observed ratio $^{123}(T_{1}^{-1})/^{121}(T_{1}^{-1})$ is given by,
\begin{equation}
\begin{split}
\label{eq4}
\frac{^{123}(T_{1}^{-1})}{^{121}(T_{1}^{-1})}=
\frac{^{123}R}{^{121}R}&\sim\frac{^{123}R_{\rm M}+^{123}R_{\rm Q}}{^{121}R_{\rm M}+^{121}R_{\rm Q}}
\end{split}
\end{equation}
Hence $^{121}R_{\rm Q}(\equiv \!\!\!^{121}(1/T_{1}T)_{\rm Q}$) is tentatively extracted from the relation expressed as, 
\begin{equation}
\begin{split}
^{121}R_{\rm Q}&\sim\ ^{121}R\left[1- \frac{1-C_Q(^{121}R/^{123}R)}{(C_{\rm M}-C_{\rm Q})(^{121}R/^{123}R)}\right]
\end{split}
\end{equation}
where the $C_{\rm M}$=0.293 and $C_{\rm Q}$=0.691, and the ratio ($^{123}R$/$^{121}R$)  obtained in Fig. \ref{Sb_T1_ref}(d). 

\begin{figure}[htbp]
\centering
\includegraphics[width=9cm]{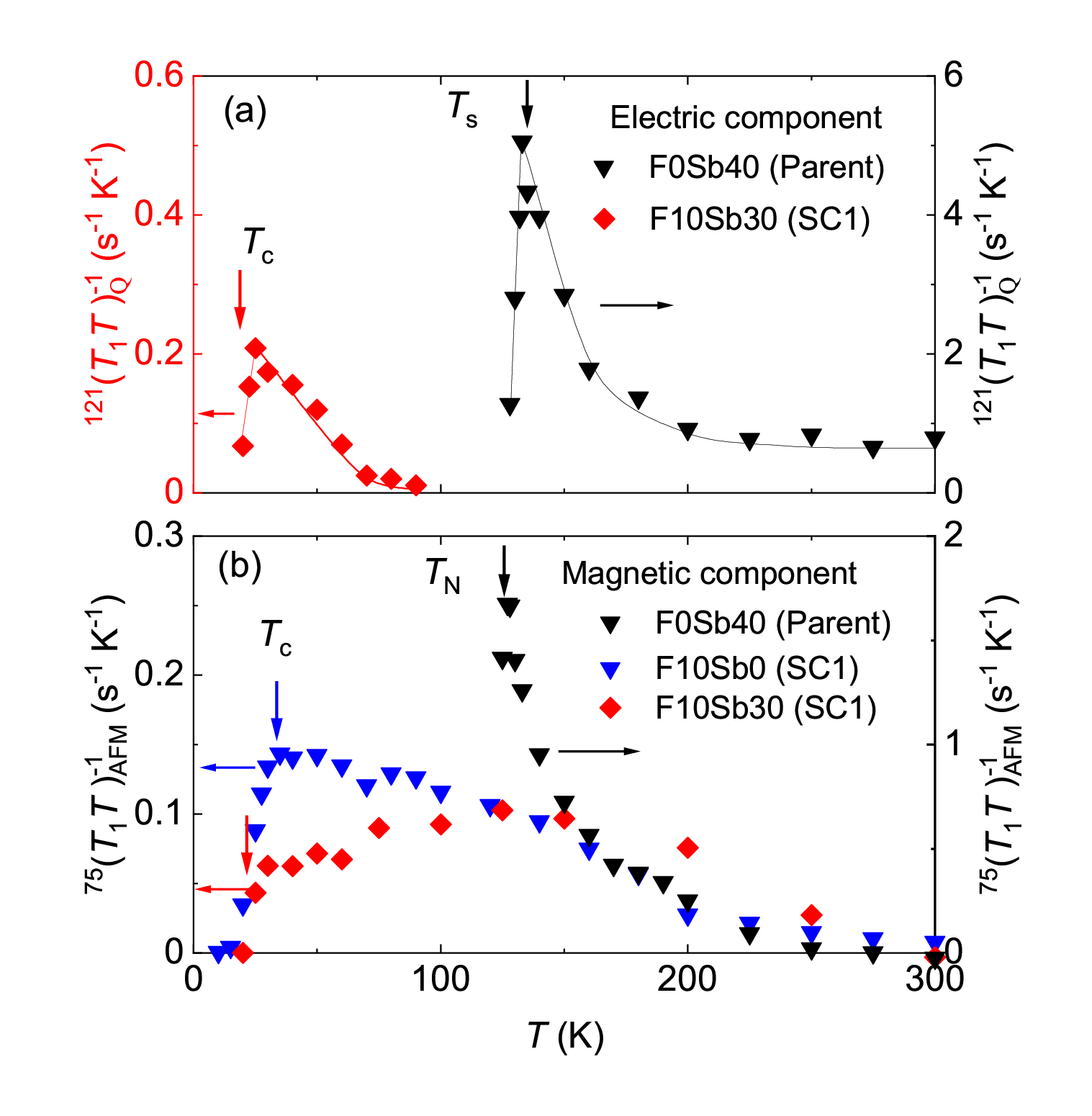}
\caption[]{(Color online)  
(a)$T$ dependence of the electric relaxation $^{121}(1/T_{1}T)_{\rm Q}$ extracted for the parent (F0Sb40) and lightly electron-doped SC state (F10Sb30), suggesting that the nematic fluctuations observed in the parent is largely suppressed but remains below 80 K in the SC1 phase even in the static $C_4$ symmetry.
(b) Spin fluctuations $^{75}(1/T_{1}T)_{\rm AFM}$ deduced for the parent (F0Sb40), and SC1(F10Sb0 and F10Sb30). 
In the lightly electron-doped SC1 region, the spin fluctuations are enhanced upon cooling, whereas the suppression of AFMSFs upon cooling becomes characteristic in Sb-doped sample(F10Sb30) that appears when pnictogen height is high\cite{T.Kouchi2022}.
}
\label{ratio_Q}
\end{figure}

Figure \ref{ratio_Q}(a) shows the outline of $T$ dependence of $^{121}(1/T_{1}T)_{\rm Q}$ component.
It enables us to estimate the $T$ evolution of the dynamical nematic fluctuations $\chi_{\rm nem}^{\prime\prime}$ at low energies that increases significantly toward $T_s$. 
In most of the previous NMR studies using the nuclear quadrupole interaction such as $^{75}$As-NMR,  the evolution of the $static$ nematicity has been discussed, which is evaluated by the static imbalance population between $p_x$ and $p_y$ orbitals at As (or Se) site for the single crystals such as 122, 11, and 111-based compounds\cite{S.H.Baek2014,T.Iye2015,R.Zhou2016,M.Toyoda2018}. 
The dynamics of nematic fluctuations in these compounds were discussed from the anisotropy of the relaxation rate of their single crystals\cite{S.H.beak2018,J.Li2020,R.Zhou2020}.
Less study has focused on the dynamics of nematic states in La1111-based compounds in the earlier $^{75}$As-NMR works due to poor sensitivity to the orbital degrees of freedom\cite{Y.Nakai2008,N.Terasaki2009,H.J.Grafe2008}.
We note that the novelty of this study is the ability to directly probe the charge fluctuations associated with nematic order, even if it is not limited to single crystal.
The current $^{121,123}$Sb nuclear probe possessing the large electric quadrupole moment ($^{121,123}Q$)  gives us an unique opportunity to extract the dynamical feature of the orbital fluctuation sensitively.

\subsection{Lightly electron-doped SC compound LaFe(As$_{0.7}$Sb$_{0.3}$)(O$_{0.9}$F$_{0.1}$) (``F10Sb30'') }

Next, the same experimental method is applied for the lightly electron-doped SC compound LaFe(As$_{0.7}$Sb$_{0.3}$)(O$_{0.9}$F$_{0.1}$) (``F10Sb30'')  ($T_c \sim$ 20 K).  
Figures \ref{Sb_T1_ref}(e)-\ref{Sb_T1_ref}(g) show the $T$ dependence of $1/T_{1}T$ measured at (e) La, (f) As, and (g) Sb sites. 
There is neither static magnetic order nor the nematic order, i.e., the tetragonal($C_4$) structure is stable in the whole $T$ range, and hence the $1/T_{1}T$ does not exhibit any peaks.  
The $T$ dependence of $1/T_{1}T$ exhibits  gradual decreases upon cooling at all the nuclear sites, which is generally seen in lightly electron-doped Fe-pnictides, ensuring the lightly electron-doped state. 
Figure \ref{Sb_T1_ref} (h) shows the $T$ dependence of their isotopic ratio $^{123}(T_{1}^{-1})/^{121}(T_{1}^{-1})$ at $^{121,123}$Sb site. 
We found that it is close to the magnetic ratio ($C_{\rm M}$ = 0.293) at high $T$, but it approaches gradually upon cooling toward the electric ratio($C_{\rm Q}$ = 0.691),  indicating that the electrical relaxation due to the dynamical nematic fluctuations between $C_4$ and $C_2$ symmetries is not negligible below 80 K for the lightly electron-doped SC1 states, although it keeps the static $C_4$ symmetry from the macroscopic point of view.
Note that the scales of $^{121}(1/T_{1}T)_{\rm Q}$ for parent (F0Sb40) and SC1(F10Sb30)  are about 10 times different, as seen in the right and left axes in Fig. \ref{ratio_Q}(a), respectively. This large difference seems to be remarkable beyond some ambiguity in the
absolute value of $T_{1}$ in this analysis.
In the previous $^{75}$As-NMR measurement of BaFe$_{2}$(As$_{0.67}$P$_{0.33}$)$_{2}$($T_{c} = 31$ K), it was revealed that the nematic fluctuations were also enhanced toward low $T$ in the superconducting phase with $C_4$ symmetry, by comparing the relaxation rate between $^{31}$P$(I = 1/2)$ and $^{75}$As$(I = 3/2)$ nucleus\cite{A.P Dioguardi2016}. The dynamics of nematic states seen in BaFe$_{2}$(As,P)$_{2}$\cite{A.P Dioguardi2016} can be observed in La1111-based compound(F10Sb30) as well. The nematic fluctuations are enhanced below 100 K in the slightly doped SC phase of both BaFe$_{2}$(As$_{0.67}$P$_{0.33}$)$_{2}$($T_{c} = 31$ K) and La1111(F10Sb30)($T_{c}=20$ K) within the $C_4$ symmetry.
From these experimental facts, the contribution of nematic fluctuations to SC may not be negligible in the iron pnictides with well nested or moderate nested FSs.



\subsection{Comparison with AFM spin fluctuations}

Next we discuss the AFM spin fluctuations (AFMSFs) probed by $^{75}$As-NMR in these compounds, since a number of the previous $^{75}$As-NMR reports indicates the observed $^{75}(1/T_1T)$ is generally dominated by magnetic relaxation, that is, $^{75}(1/T_1T)_{\rm M}$\cite{T.Imai2009,Ning2010,Y.Nakai2010,T.Oka2012,T.Shiota2016,P.Wiecki2018,T.Kouchi2022}.
According to these previous works, it is known that the observed $^{75}(1/T_1T)_{\rm M}$ can be decomposed as $(1/T_1T)_{\rm AFM}+(1/T_1T)_0$, where the first term $(1/T_1T)_{\rm AFM}$ represents the component of AFMSFs at finite wave vector $\vec q$ = $Q$ in Eq.(1), and the second term $(1/T_1T)_0$ is the component related to the square of the density of states ($N(E_{\rm F})$) that is proportional to the spin part of Knight shift ($K_{\rm s}$).
The detail of this method is described elsewhere\cite{T.Kouchi2022} and in supplemental information\cite{sup}.

As a result, Fig. \ref{ratio_Q}(b) shows the estimated $^{75}(1/T_{1}T)_{\rm AFM}$ for the parent(F0Sb40) and SC1(F10Sb0 and F10Sb30).
To evaluate the substitution effect of Sb in the SC1 phase, the result of Sb-free compound F10Sb0($T_c$ = 28 K)  in the same doping level is compared.
The $^{75}(1/T_{1}T)_{\rm AFM}$  in the SC1(F10Sb0 and F10Sb30) phase  increases upon cooling, suggesting the presence of the low-energy spin fluctuations that can be attributed to the well-nested FSs mostly composed of two $d_{xz/yz}$ orbitals\cite{T.Shiota2016}.
We note that this is not the case of the simple Curie-Weiss type behavior, namely, the $^{75}(1/T_{1}T)_{\rm AFM}$ for Sb-substituted F10Sb30 is more suppressed below 100 K than for Sb-free F10Sb0.
This behavior is more prominent when the energy level of $d_{xy}$ orbital band approaches to the Fermi level, when the pnictogen height becomes high, i.e. Sb content increases\cite{T.Kouchi2022}.
Thus, such small suppression of AFMSFs upon cooling is accounted for by the increment of $d_{xy}$ orbital band contributions in addition to the predominant $d_{xz/yz}$ bands\cite{K.Kuroki2008,T.Kawashima2021}.
According to the spin-based scenario\cite{I.I.Mazin2008,K.Kuroki2008}, the presence of the low-energy spin fluctuations are one of the important factors for the increment of $T_c$ in the lightly doped SC1 phase, which is consistent with the present experimental fact that it is more significant in Sb-free F10Sb0($T_c$ = 28  K) than in F10Sb30($T_c$ = 20K) since these locate in the lightly doped SC1 region.
This is in contrast to the heavily electron-doped SC3 region of La1111(See Fig. 1), where the lack of spin fluctuations at low energies is more significant and rather favorable for SC: In other word, the presence of the finite-energy spin fluctuations originating from the predominant  $d_{xy}$ band with strong correlations (less contribution from $d_{xz/yz}$ bands) is one of the key factors for the appearance of reemergent SC3 phase in the case of heavily electron-doped regime\cite{T.Kouchi2022,K.Suzuki2014,H.Arai2015,M.Nakata2017,K.Matsumoto2020}.

Consequently, in these experiments, we succeeded in the detection of both nematic and spin fluctuations in the SC1 phase, as presented in Fig. 4(a) and 4(b), respectively.
It implies that the spin and orbital degrees of freedom are inseparable in nature.
To identify the qualitative roles of the effect of the orbital degrees of freedom over wider doping region as a function of $T_c$ values, the future systematic Sb-NMR measurements are desired from lightly doped SC1 to heavily doped SC3 phases.

\section{Summary}

In summary, we succeeded in detecting the dynamics of the stripe-type orbital fluctuations in the Sb-substituted La1111 compounds by  the isotope $^{121,123}$Sb-NMR probes with the large $^{121,123}$Sb quadrupole moments.
We revealed that the nematic fluctuations of the stripe-type alignment of Fe-$3d_{xz}$(or $3d_{yz}$) orbital are critically enhanced toward $T_{s}\sim135$ K slightly higher than  $T_{\rm N}\sim$ 125 K in the parent(F0Sb40).
In the lightly electron-doped SC1 phase, such nematic  fluctuations are largely suppressed  but remains a small enhancement below 80 K down to the $T_c$, in spite of keeping the tetragonal($C_4$) symmetry.
As for the spin degrees of freedom, the AFMSFs are also observed on the SC1 phase due to the well-nested FSs derived mainly from $3d_{xz/yz}$ orbitals, with some additional contribution from $3d_{xy}$ orbital especially when the pnictogen height is large.
The results suggest that both the spin and orbital degrees of freedom on the $3d_{xz/yz}$ orbitals can be seen in lightly electron-doped SC1 phase, implying that they are not completely separated in the Fe-pnictide superconductors.
The future systematic NMR measurements over wide electron doping region will give us a qualitative evaluation of the effect of the orbital degrees of freedom from SC1 to heavily electron-doped SC3.
We emphasize here that the present isotope Sb-NMR provides with a new approach to investigate the correlation between orbital/spin fluctuations and SC state in La1111 system systematically, even in polycrystalline samples.




\vskip\baselineskip
{\footnotesize 
One of the author (T.K.) is supported by a JSPS Fellowship (Grant No. 21J14053). 
This work was supported by JSPS KAKENHI (Grant No. 18K18734),  Iketani science and technology foundation, Izumi science and technology foundation, Casio science promotion foundation, and Takahashi Industrial and Economic Research Foundation.
}


\end{document}


\title{Dynamics of orbital degrees of freedom probed via isotope $^{121,123}$Sb nuclear quadrupole moments in Sb-substituted iron-pnictide superconductors}

\author{T. Kouchi}\email[]{e-mail  address: takayoshi.kouchi@rs.tus.ac.jp}
\affiliation{Graduate School of Engineering Science, Osaka University, Osaka 560-8531, Japan}
\affiliation{Department of Applied Physics, Tokyo University of Science, Tokyo 125-8585, Japan}
\author{K. Yoshinaga}
\author{T. Asano}
\affiliation{Graduate School of Engineering Science, Osaka University, Osaka 560-8531, Japan}
\author{S. Nishioka}
\affiliation{Graduate School of Engineering Science, Osaka University, Osaka 560-8531, Japan}
\affiliation{National Institute for Materials Science, Ibaraki 305-0003, Japan}
\author{M. Yashima}
\author{H. Mukuda}\email[]{e-mail  address: mukuda.hidekazu.es@osaka-u.ac.jp}
\affiliation{Graduate School of Engineering Science, Osaka University, Osaka 560-8531, Japan}
\author{A. Iyo}
\affiliation{National Institute of Advanced Industrial Science and Technology (AIST), Ibaraki 305-8568, Japan}
\author{T. Kawashima}
\author{S. Miyasaka}
\affiliation{Graduate School of Science, Osaka University, Osaka 560-0043, Japan}

\date{\today}

\maketitle
\section*{Supplemental information}

\subsection*{Evaluation of the component of the spin fluctuations}
\begin{figure}[htbp]
\centering
\includegraphics[width=12cm]{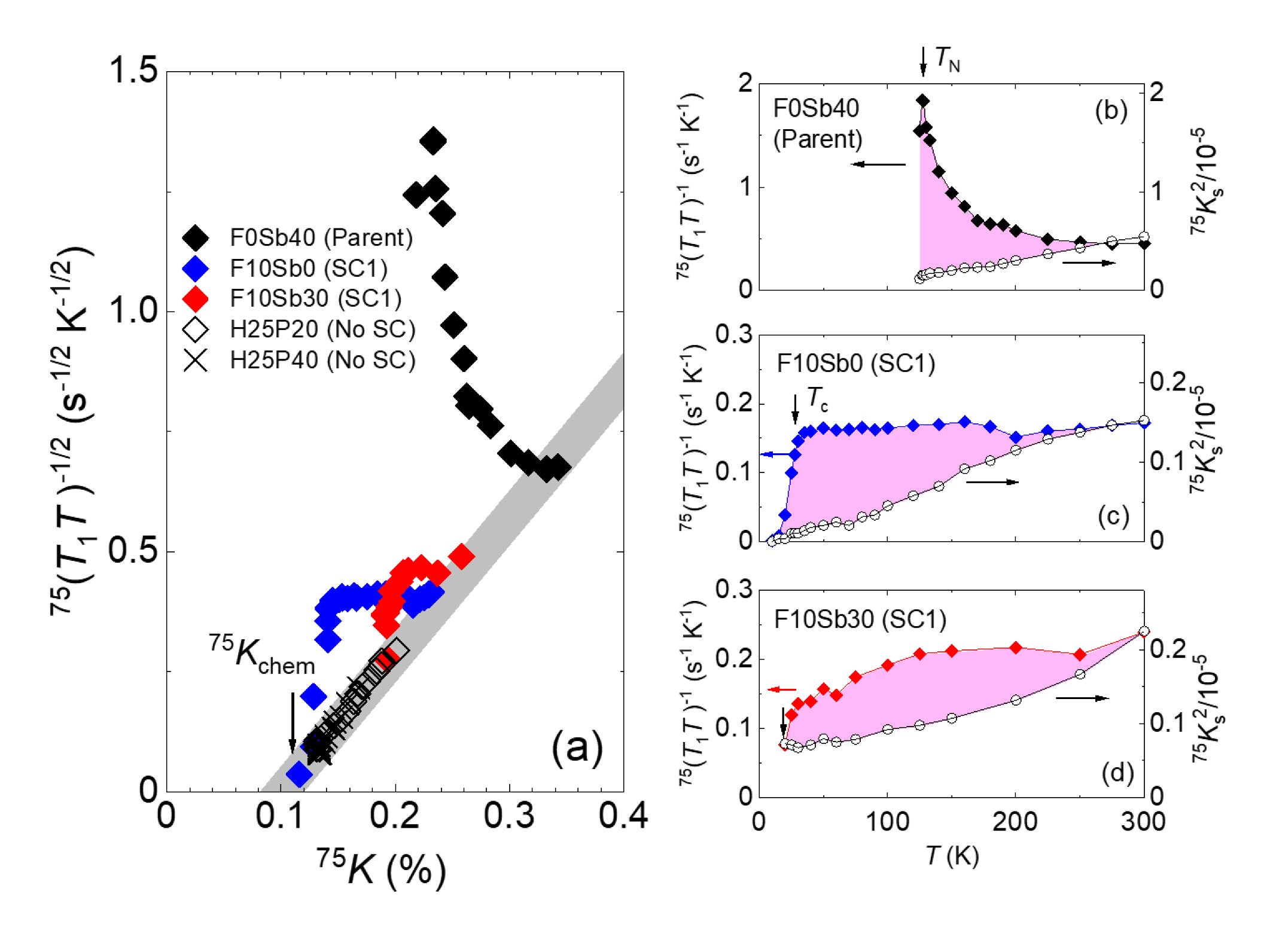}
\caption[]{(Color online)  
(a) Plot of $^{75}$$(1/T_{1}T)^{0.5}$ vs $^{75}$$K$ for the parent (``F0Sb40''), the lightly electron-doped SC1 phases (``F10Sb0'' and ``F10Sb30'). 
The gray line represents the results for the non SC samples with no spin fluctuations (``H25P20'' and ``H25P40'') previously reported\cite{T.Kouchi2022}. 
(b)-(d) $T$ dependence of $^{75}$$(1/T_{1}T)$ and $K_{s}^{2}$. 
The hatched regions correspond to the component of $(1/T_{1}T)_{\rm AFM}$, which corresponds to Fig. 4(b) in the main text. 
}
\label{supplement}
\end{figure}

According to the previous $^{75}$As NMR reports, the observed $^{75}(1/T_1T)$ is known to be dominated by magnetic relaxation, i.e. $^{75}(1/T_1T)\sim^{75}(1/T_1T)_{\rm M})$.
It can be expressed as $(1/T_1T)_{\rm AFM}+(1/T_1T)_0$\cite{T.Imai2009,Ning2010,Y.Nakai2010,T.Oka2012,T.Shiota2016,P.Wiecki2018,T.Kouchi2022}. 
The first term $(1/T_1T)_{\rm AFM}$ is the component of AFMSFs, and the second term $(1/T_1T)_0$ that is proportional to the spin part of Knight shift ($K_{\rm s}$). 
The $K_{\rm s}$ is obtained by subtracting the chemical shift $K_{\rm chem}$ from the observed Knight shift $K$. 
The $K_{\rm chem}$ is estimated by the plot of $^{75}$$(1/T_{1}T)^{0.5}$ vs $^{75}$$K$ shown in Fig. \ref{supplement}(a).
If the spin fluctuations are absent, the linear relation is expected, as shown by the gray line in the figure, which corresponds to Korringa relation seen in the non-correlated normal metals. 
Actually, the results for the non-correlated Fe-pnictides with no spin fluctuations (``H25P20'' and ``H25P40'') locate on the line, as previously reported\cite{T.Kouchi2022}. 
Here ``H25P20'' and ``H25P40'' represents non-superconducting LaFe(As$_{0.8}$P$_{0.2}$)(O$_{0.75}$H$_{0.25}$) and  LaFe(As$_{0.6}$P$_{0.4}$)(O$_{0.75}$H$_{0.25}$), respectively\cite{T.Kouchi2022}.
This plot gives us an estimation of $K_{\rm chem}$ to be $\sim$ 0.11\%. 
The data for the parent (``F0Sb40''), SC1 phases (``F10Sb0'' and ``F10Sb30'') deviates from the line at low temperatures, whereas the data approach to this line at high temperatures, indicating that the spin fluctuations appear upon cooling. 
Figures \ref{supplement}(b)-(d) show the $T$ dependence of $^{75}(1/T_{1}T)$ and $K^{2}_{s}$ for the parent (``F0Sb40''), and the SC1 phases (``F10Sb0'' and ``F10Sb30'').
The component of $(1/T_{1}T)_{\rm AFM}$  is obtained by subtracting the second term $(1/T_1T)_0$ from the observed $1/T_1T$, as shown by the hatched area in the figure, which corresponds to Fig. 4(b) in the main text. 

